\documentclass[aps,pra,showpacs,twocolumn,groupedaddress]{revtex4}
\usepackage{graphicx,amsmath,amssymb}
\begin{document}
\newtheorem{Theorem}{Theorem}
\newtheorem{Lemma}{Lemma}

\title{Majorization in Quantum Adiabatic Algorithms}

\author{Zhaohui Wei}
\email{weich03@mails.tsinghua.edu.cn }
\author{Zhengfeng Ji}
\email{jizhengfeng98@mails.tsinghua.edu.cn }
\author{Mingsheng Ying}
\email{yingmsh@mail.tsinghua.edu.cn }

\affiliation{ State Key Laboratory of Intelligent Technology and
Systems, Department of Computer Science and Technology, Tsinghua
University, Beijing, China, 100084}

\begin{abstract}

The majorization theory has been applied to analyze the mathematical
structure of quantum algorithms. An empirical conclusion by
numerical simulations obtained in the previous literature indicates
that step-by-step majorization seems to appear universally in
quantum adiabatic algorithms. In this paper, a rigorous analysis of
the majorization arrow in a special class of quantum adiabatic
algorithms is carried out. In particular, we prove that for any
adiabatic algorithm of this class, step-by-step majorization of the
ground state holds exactly. For the actual state, we show that
step-by-step majorization holds approximately, and furthermore that
the longer the running time of the algorithm, the better the
approximation.

\end{abstract}

\maketitle

\section{INTRODUCTION}

In the past two decades, quantum computation has attracted a great
deal of attention, for it was demonstrated that the performance of
quantum algorithms exceeds that of all known classical corresponding
algorithms for some computational tasks. Among all quantum
algorithms proposed so far, Shor's factorization algorithm
\cite{SHOR94} and Grover's search algorithm \cite{GROVER97} are two
famous examples. However, the design of quantum algorithms seems to
be very difficult \cite{Shor03}. Therefore, uncovering some
underlying mathematical structure of quantum algorithms becomes a
very important question. For example, it has been observed that
majorization theory seems to play an important role in the
efficiency of quantum algorithms \cite{OLM021,OLM022,LM02}. The
intuition is that in many quantum algorithms, the initial state of
the system is an equal superposition state and the final state
before measurement is some computational basis state corresponding
to the final result. In the process of computation, the probability
distribution associated to the state of the system in the
computational basis is step-by-step majorized until it is maximally
ordered. In \cite{OLM021}, by carrying out a systematic analysis of
a wide variety of quantum algorithms from the majorization theory
point of view, R. Or\'{u}s et al. concluded that step-by-step
majorization is found in the known instances of fast and efficient
algorithms, such as quantum fourier transform, Grover's algorithm,
the algorithm for the hidden affine function problem. On the other
hand, R. Or\'{u}s et al. offered an example to show that some
quantum algorithms, which do not give any computational speed-up,
violates step-by-step majorization. These facts indicate that
step-by-step majorization seems to be necessary for the efficiency
of quantum algorithms.

In \cite{OLM021} and \cite{LM02}, the analysis of the role of
majorization in quantum adiabatic algorithms, a novel paradigm for
the design of quantum algorithms, was also carried out. Through
numerical simulations to several special cases R. Or\'{u}s et al.
got an empirical conclusion that quantum algorithms based on
adiabatic evolution naturally show step-by-step majorization
provided that the Hamiltonians and the initial state are chosen with
sufficient symmetry and the evolution is slow enough.

In a quantum adiabatic algorithm, the evolution of the quantum
register is governed by a hamiltonian that varies continuously and
slowly. If the initial state of the system is the ground state of
the initial hamiltonian, the state of the system at any moment in
the whole process of computation will differ from the ground state
of the hamiltonian at that moment by a negligible amount. Thus, in a
quantum adiabatic algorithm the ground state of the hamiltonian is a
``guide'', and the actual state of the system always evolves around
this guide. In this paper, we will analyze the majorization arrow in
a special class of quantum adiabatic algorithms. We prove that in
any algorithm of this class step-by-step majorization of the ground
state holds perfectly. For the actual state, we show that
step-by-step majorization holds approximately and that the longer
the running time, the better the approximation. Thus the results
obtained in this paper offers stronger evidences to support the
conclusion drawn by R. Or\'{u}s et al.

The rest of the paper is organized as follows. In Sec. II we briefly
review quantum adiabatic computation and majorization theory. In
Sec. III we prove that step-by-step majorization of the ground state
holds. In Sec. IV we discuss step-by-step majorization of the actual
state. Finally, in Sec. V we summarize our conclusions and discuss
the role majorization plays in the efficiency of quantum algorithms.

\section{PRELIMINARIES}

For the convenience of the readers, in this section we will recall
quantum adiabatic computation and majorization theory.

Quantum adiabatic computation, proposed by Farhi \cite{FGGS00}, is
based on quantum adiabatic evolution. Suppose the state of a quantum
system is $|\psi(t)\rangle(0\leq t\leq T)$, which evolves according
to the Schr\"{o}dinger equation
\begin{equation}
i\frac{d}{dt}|\psi(t)\rangle=H(t)|\psi(t)\rangle,
\end{equation}
where $H(t)$ is the Hamiltonian of the system. Suppose $H_0=H(0)$
and $H_1=H(T)$ are the initial and the final Hamiltonians of the
system. Then we let the hamiltonian of the system vary from $H_0$ to
$H_1$ slowly along some path. For example, an interpolation path is
one choice,
\begin{equation}
H(t)=f(t)H_0+g(t)H_1,
\end{equation}
where $f(t)$ and $g(t)$ are continuous functions with $f(0)=g(T)=1 $
and $f(T)=g(0)=0$ ($T$ is the running time of the evolution). Let
$|E_0,t\rangle$ and $|E_1,t\rangle$ be the ground state and the
first excited state of the Hamiltonian at time t, and let $E_0(t)$
and $E_1(t)$ be the corresponding eigenvalues. The quantum adiabatic
theorem \cite{LIS55} shows that we have
\begin{equation}
|\langle E_0,T|\psi(T)\rangle|^{2}\geq1-\varepsilon^2,
\end{equation}
provided that
\begin{equation}
\frac{D_{max}}{g_{min}^2}\leq\varepsilon,\ \ \ \ 0<\varepsilon\ll1,
\end{equation}
where $g_{min}$ is the minimum gap between $E_0(t)$ and $E_1(t)$
\begin{equation} g_{min}=\min_{0\leq t \leq T}[E_1(t)-E_0(t)],
\end{equation}
and $D_{max}$ is a measurement of the evolving rate of the
Hamiltonian
\begin{equation}
D_{max}=\max_{0\leq t \leq T}|\langle
E_1,t|\frac{dH}{dt}|E_0,t\rangle|.
\end{equation}

Quantum adiabatic computation is a novel paradigm for the design of
quantum algorithms. For example, Quantum search algorithm proposed
by Grover \cite{GROVER97} has been implemented by quantum adiabatic
computation in \cite{RC02}. Recently, the new paradigm for quantum
computation has been used to try to solve some other interesting and
important problems, such as Deutsch-Jozsa problem
\cite{DKK02,SL05,WY06}, hidden subgroup problem \cite{RAO03}, 3SAT
problem \cite{FGGS00,ZH06}, traveling salesman problem \cite{TDK06}
and Hilbert's tenth problem \cite{TDK01}.

Let's first define a special class of quantum adiabatic algorithms,
on which we will focus in this work. Suppose $f:\{0,1\}^n\rightarrow
R$ is a function bounded by a polynomial of n. Let $H_0$ and $H_1$
be the initial and the final hamiltonians of a quantum adiabatic
evolution with a linear path $H(t)$. Concretely,
\begin{equation}
H_0=I-|\alpha\rangle\langle\alpha|,
\end{equation}
\begin{equation}
H_1=\sum\limits_{i=1}^{N}{f(i)|i\rangle\langle i|},
\end{equation}
\begin{equation}
H(s)=(1-s)H_0+sH_1,
\end{equation}
where
\begin{equation}
|\alpha\rangle=\frac{1}{\sqrt{N}}\sum\limits_{i=1}^{N}{|i\rangle}, \
\ N=2^n,
\end{equation}
and $s=s(t)$ a continuous increasing function with $s(0)=0$ and
$s(T)=1$ ($T$ is the running time of the quantum adiabatic
evolution). According to quantum adiabatic theorem, this class of
algorithms can be used to minimize the function $f(i), i=1,2,...,N$.
The quantum adiabatic algorithms for search problem in \cite{RC02},
hidden subgroup problem in \cite{RAO03}, 3SAT problem in \cite{ZH06}
and traveling salesman problem in \cite{TDK06} belong to this class.

Now let's turn to the majorization theory. Majorization is an
ordering on N-dimensional real vectors. Suppose
$x=(x_1,x_2,...,x_N)$ and $y=(y_1,y_2,...,y_N)$ are two
N-dimensional vectors. If $x$ is majorized by $y$, $x$ is more
disordered than another. To be concrete, let $x^\downarrow$ mean $x$
re-ordered so the components are in decreasing order. We say $x$ is
majorized by $y$, namely $x\prec y$, provided
$\sum\limits_{i=1}^{k}{x_i^\downarrow}\leq
\sum\limits_{i=1}^{k}{y_i^\downarrow}$ for $k=1,2,...,N-1$ and
$\sum\limits_{i=1}^{N}{x_i^\downarrow}=\sum\limits_{i=1}^{N}{y_i^\downarrow}$.
It has been proven that majorization is at the heart of the solution
of a large number of quantum information problems. For example,
majorization characterizes when one quantum bipartite pure states
can be transformed to another deterministically via local operations
and classical communication \cite{Nilsen99}. For more details about
majorization, see \cite{RB97}.

In \cite{OLM021} and \cite{LM02}, majorization theory was related to
quantum algorithms. It can be shown as follows: let
$|\psi^{(m)}\rangle$ be the state of the register of a quantum
computers at an operating stage labeled by $m=1,...M$, where $M$ is
the total number of steps in the algorithm. Let $N$ be the dimension
of the Hilbert space. Suppose $\{|i\rangle\}_{i=1}^N$ is the basis
in which the final measurement is performed. Then suppose in this
basis the state $|\psi^{(m)}\rangle$ is
\begin{equation}
|\psi^{(m)}\rangle=\sum\limits_{i=1}^{N}{a_i^{(m)}|i\rangle}.
\end{equation}
If we measure $|\psi^{(m)}\rangle$ in the basis
$\{|i\rangle\}_{i=1}^N$, the probability distribution associated to
this state is $p^{(m)}=\{p_i^{(m)}\}$, where
\begin{equation}
p_i^{(m)}\equiv |a_i^{(m)}|^2=|\langle i|\psi^{(m)}\rangle|^2, \ \
i=1,2,...,N.
\end{equation}
If $p^{(m)} \prec p^{(m+1)}$ for every $m$, we say this algorithm
enjoys the majorization relation step by step.

Especially, majorization theory has been applied to analyze quantum
adiabatic algorithms. Suppose $t_1$ and $t_2$ are two arbitrary time
point in an adiabatic evolution, and $t_1<t_2$. If it always holds
that the probability distribution associated to the state of the
system at $t_1$ is majorized by that at $t_2$, we say this adiabatic
algorithm enjoys step-by-step majorization. In \cite{OLM021} and
\cite{LM02}, R. Or\'{u}s et al. studied majorization in local and
global quantum adiabatic search algorithms. Note that both these two
algorithms belong to the class of quantum adiabatic algorithms we
will discuss.

\section{STEP-BY-STEP MAJORIZATION OF THE GROUND STATE}

As mentioned above, in a quantum adiabatic algorithm, the state of
the system at any time is always close to the ground state of the
hamiltonian of that moment with a small distance. So analyzing the
evolution of the ground state may help us to understand that of the
actual state.

In this section, we prove that, for any quantum adiabatic algorithm
of the class of quantum adiabatic algorithms described by
Eqs.(7)-(10), step-by-step majorization of the ground state holds
perfectly. Before proving this result, we first consider the
following two lemmas.

We have known that the purpose of quantum adiabatic algorithms given
by Eqs.(7)-(10) is to find the minimum of the function $f(x)$. The
following lemma shows that only the range of $f(x)$ affects our
discussion and the distribution of this set does not.

\begin{Lemma} Suppose there are two quantum adiabatic evolutions given by Eqs.(7)-(10) with different $H_1$. Concretely, these two final
hamiltonians are
\begin{equation}
H_1=\sum\limits_{i=1}^{N}{f(i)|i\rangle\langle i|}
\end{equation}
and
\begin{equation}
H_1'=\sum\limits_{i=1}^{N}{f'(i)|i\rangle\langle i|},
\end{equation}
where $f'(i)=f(\pi(i))$. Here $\pi$ is a permutation of $1,2,...,N$.
Let the ground states of these two quantum adiabatic evolution be
real vectors $(a_1,a_2,...,a_N)^T$ and $(a_1',a_2',...,a_N')^T$,
respectively. Then we have
\begin{equation}
a_i'=a_{\pi(i)}, 1\leq i \leq N.
\end{equation}
\end{Lemma}

{\it Proof.} The proof is easy as long as we note that
$P_{\pi}H'(s)P_{\pi}=H(s)$, where $P_{\pi}$ is a permutation matrix
such that $P_{\pi}|i\rangle=|\pi(i)\rangle$ and
\begin{equation}
H'(s)=(1-s)H_0+sH_1'.
\end{equation}\hfill $\Box$

It can be shown that we can choose the global phase of the ground
state of $H(s)$ given by Eq.(9) such that it is a real vector. In
this paper, we always assume ground states to be real.

Usually, it is difficult to work out the ground state of $H(s)$
exactly. However, the following lemma indicates that there are close
relations among the components of this ground state. Our proof for
the main result of this section is based on these relations.

\begin{Lemma} Suppose there is a quantum adiabatic evolutions given by Eqs.(7)-(10). Let real vector
\begin{equation}
|\psi(s)\rangle=(a_1,a_2,...,a_N)^T
\end{equation}
be the ground state of this quantum adiabatic evolution and
$\lambda(s)$ the corresponding eigenvalue. Then we have
\begin{equation}
(t(s)+sf(i))a_i = (t(s)+sf(j))a_j, \ \  i,j=1,2,...,N,
\end{equation}
where $t(s)=1-s-\lambda(s)$ is a strictly decreasing function of
$s$.
\end{Lemma}

{\it Proof.} By the definitions of $|\psi(s)\rangle$ and
$\lambda(s)$, we have
\begin{equation}
H(s)|\psi(s)\rangle=\lambda(s)|\psi(s)\rangle.
\end{equation}
Substituting Eq.(9) and Eq.(17) into Eq.(19) yields
\begin{equation}
\frac{1-s}{N}\sum_{i=1}^N{a_i}=t(s)a_i+sf(i)a_i,
\end{equation}
where $t(s)=1-s-\lambda(s)$. Note that $t(s)$ is the biggest
eigenvalue of
\begin{equation}
G(s)=(1-s)I-H(s).
\end{equation}
For every $s\in[0,1)$ and $ds>0$, an explicit calculation shows that
\begin{equation}
G(s+ds)-G(s)=-ds\times(|\alpha\rangle\langle\alpha|+\sum\limits_{i=1}^{N}{f(i)|i\rangle\langle
i|}).
\end{equation}
Because $G(s)-G(s+ds)$ is a strictly positive matrix, it can be
shown that $t(s)$ is a strictly decreasing function of $s$
\cite{RB97}. It is easy to get $t(0)=1$ and $t(1)=0$. Then we have
$0<t(s)<1$ for any $s\in(0,1)$.

By Eq.(20) we can obtain
\begin{equation}
(t(s)+sf(i))a_i = (t(s)+sf(j))a_j, \ \  i,j=1,2,...,N.
\end{equation}\hfill $\Box$

Now we are able to present the main result of this section. It
establishes the step-by-step majorization property of the ground
state of $H(s)$.

\begin{Theorem} Suppose $H_0$ and $H_1$ given by Eq.(7) and Eq.(8)
are the initial and the final hamiltonians of a quantum adiabatic
algorithm. Suppose this quantum adiabatic algorithm has a linear
path given by Eq.(9). Then the ground state of this algorithm shows
perfect step-by-step majorization.

\end{Theorem}

{\it Proof.} Suppose the ground state of $H(s)$ is
\begin{equation}
|\psi(s)\rangle=(a_1,a_2,...,a_N)^T,
\end{equation}
and the corresponding eigenvalue is $\lambda(s)$. Suppose
$\min_{1\leq i \leq N}f(i)=0$. Otherwise we can let
\begin{equation}
H(s)=H(s)-s\times I\times\min_{1\leq i \leq N}f(i),
\end{equation}
which doesn't change the ground state of $H(s)$. For convenience we
suppose $f(1)\leq f(2)\leq f(3)\leq ... \leq f(N)$, which doesn't
affect our analysis for majorization later by Lemma 1. On the other
hand, by Lemma 2 we have
\begin{equation}
(t(s)+sf(i))a_i = (t(s)+sf(j))a_j, \ \  i,j=1,2,...,N,
\end{equation}
where $t(s)$ is defined as before. Substituting Eq.(26) into
\begin{equation}
\sum_{i=1}^N{{a_i}^2}=1
\end{equation}
gives
\begin{equation}
{a_1}^2=\frac{1}{\sum_{i=1}^N{(\frac{1}{1+\frac{s}{t(s)}f(i)})^2}}.
\end{equation}
Note that $t(s)$ is a strictly decreasing function of $s$, which
means $a_1$ is a strictly increasing function of $s$. For any other
$a_i$, the monotony is a little more complicated. It's possible that
they are not monotonous. However, we can prove that their increasing
and decreasing are well-regulated. Concretely, for $s>0$ and
$1-s\geq ds>0$, let $({a_1}',{a_2}',...,{a_N}')^T$ be the ground
state of $H(s+ds)$. Then we have if $a_i\geq {a_i}'$, $a_j\geq
{a_j}'$, where $i<j$.

This conclusion can be proved as follows. From Eq.(26), we obtain
\begin{equation}
\frac{a_i}{a_j}=\frac{t(s)+sf(j)}{t(s)+sf(i)},
\end{equation}
and
\begin{equation}
\frac{{a_i}'}{{a_j}'}=\frac{t(s+ds)+(s+ds)f(j)}{t(s+ds)+(s+ds)f(i)}.
\end{equation}
Because $t(s)$ is a strictly decreasing function, it can be checked
that
\begin{equation}
\frac{t(s)+sf(j)}{t(s)+sf(i)}\leq\frac{t(s+ds)+(s+ds)f(j)}{t(s+ds)+(s+ds)f(i)}.
\end{equation}
So,
\begin{equation}
\frac{a_i}{a_j}\leq\frac{{a_i}'}{{a_j}'}.
\end{equation}
Thus if ${a_i}'\leq a_i$, we have ${a_j}'\leq a_j$.

According to the discussion above, we know that for every
$s\in(0,1)$ there is a special integer $i_0(s)$. When $i\leq i_0(s)$
we have $a_i\leq {a_i}'$ and when $i>i_0(s)$ we have $a_i>{a_i}'$.

Now we are in a position to prove our main conclusion. Namely,
\begin{equation}
({a_1}^2,{a_2}^2,...,{a_N}^2)^T \prec
({a_1}'^2,{a_2}'^2,...,{a_N}'^2)^T.
\end{equation}

Firstly, according to Eq.(26) it can be checked that the components
of $({a_1}^2,{a_2}^2,...,{a_N}^2)^T$ and
$({a_1}'^2,{a_2}'^2,...,{a_N}'^2)^T$ are in decreasing order.
Secondly, for any $s\in(0,1)$ and any $k=1,2,...,N$, if $k\leq
i_0(s)$, we have $\sum\limits_{i=1}^{k}{a_i^2}\leq
\sum\limits_{i=1}^{k}{a_i'^2}$ because $a_i\leq a_i'$ for $i\leq k$.
If $k>i_0(s)$, $a_i>a_i'$ for $i\geq k$, so we have
$\sum\limits_{i=k+1}^{N}{a_i^2}\geq
\sum\limits_{i=k+1}^{N}{a_i'^2}$. Thus we also get
$\sum\limits_{i=1}^{k}{a_i^2}\leq \sum\limits_{i=1}^{k}{a_i'^2}$
because
$\sum\limits_{i=1}^{N}{a_i^2}=\sum\limits_{i=1}^{N}{a_i'^2}=1$. This
completes the proof of Eq.(33). Namely, step-by-step majorization of
the guide state holds perfectly. \hfill $\Box$

Note that if the form of Eq.(7) doesn't change, we can replace the
ground state $|\alpha\rangle$ in Eq.(7) with any other vector of
Hadamard basis and get the same conclusion. Because it can be proved
if $|\alpha\rangle$ is replaced by any other vector of Hadamard
basis, for any $s$ any component of the ground state of $H(s)$ will
not change up to the sign. Moreover, it can be shown that the path
in Eq.(9) along which the hamiltonian varies can also be replaced by
any interpolation path in Eq.(2) provided $\frac{g(t)}{f(t)+g(t)}$
is a increasing function of $t$, which doesn't destroy step-by-step
majorization either.

\section{STEP-BY-STEP MAJORIZATION OF THE ACTUAL STATE}

In this section, based on the result of the above section we
consider the majorization relation in the actual state of quantum
adiabatic algorithms of the class discussed in this paper. We show
that step-by-step majorization of the actual state holds
approximately, and the degree of the approximation is determined by
the running time \cite{OLM021,LM02}.

Suppose in a quantum adiabatic evolutions given by Eqs.(7)-(10), the
actual state of the system is
\begin{equation}
|\psi'(s)\rangle=(b_1,b_2,...,b_N)^T.
\end{equation}
Let
\begin{equation}
B_k=\sum_{i=1}^k{|b_i|^2}, \ \ k=1,2,...,N.
\end{equation}
In \cite{OLM021} R. Or\'{u}s et al. studied $s-B_1$ curve ($B_1$ is
the probability of finding the right solution) and $s-B_2$ curve of
global quantum adiabatic evolution for search problem by numerical
simulations. If step-by-step majorization holds perfectly, these
curves should be monotonous. However, they observed that oscillation
appears at the end of $s-B_1$ curve and $s-B_2$ curve, which
destroys step-by-step majorization (See Figure.1). Furthermore, they
also observed that the oscillation becomes weaker and weaker and
step-by-step majorization tends to appear as long as the running
time becomes longer and longer.

\begin{figure}[htb]
\centerline{\includegraphics[angle=0,width=3.3in]{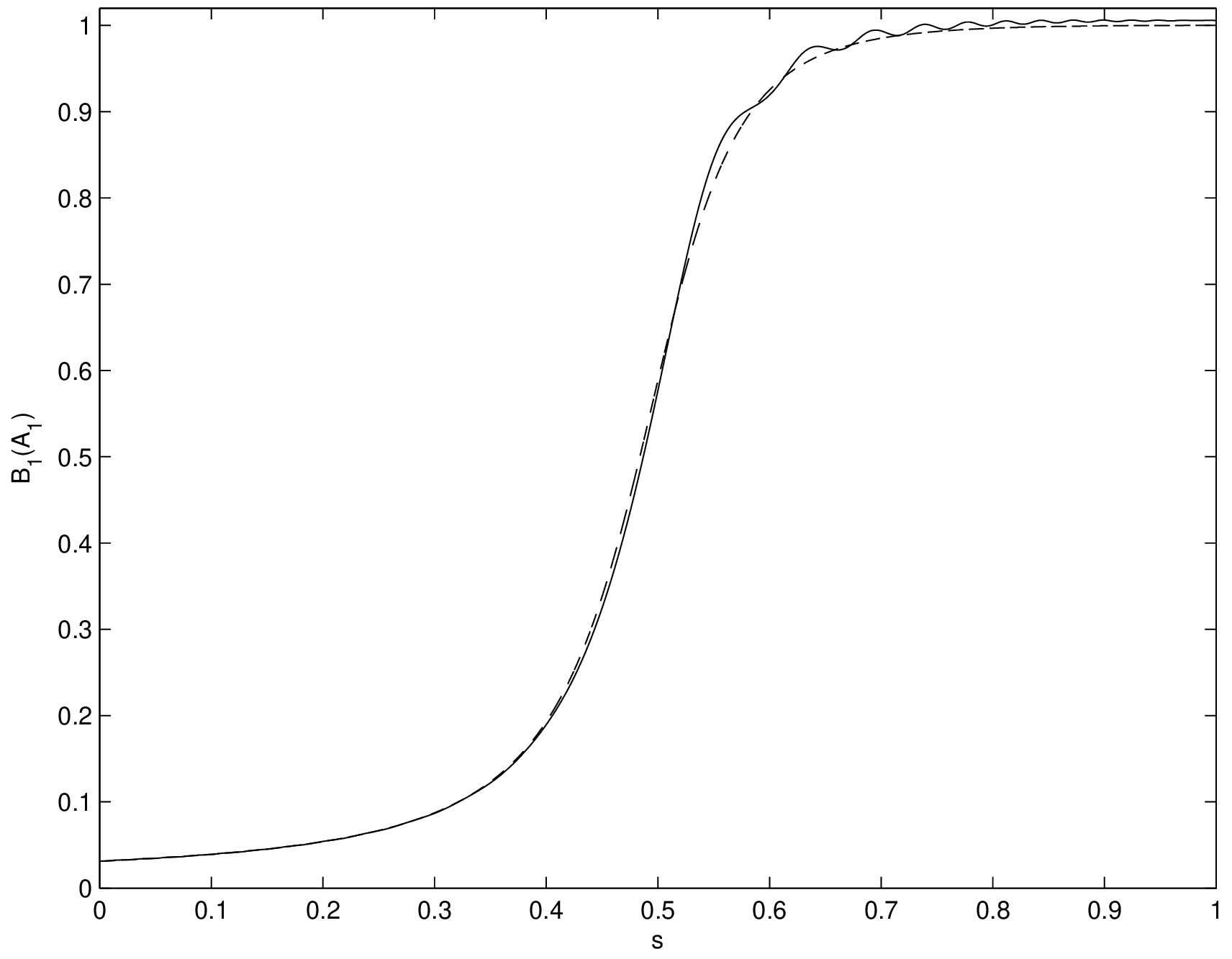}} \caption{A
case that oscillation appears at the end of $s-B_1$ curve (the solid
curve). The dashed curve is $s-A_1$ curve.} \label{fig:gaps_Id}
\end{figure}

Now, we prove that for any quantum adiabatic evolution of the class
discussed in this paper, the oscillation at the end of $s-B_k (1\leq
k \leq N)$ curve, if any, will continue decreasing in amplitude if
the running time becomes longer and longer.

We consider an arbitrary state of the system near the end of the
quantum adiabatic evolution. Let
\begin{equation}
A_k=\sum_{i=1}^k{a_i^2}, \ \ k=1,2,...,N,
\end{equation}
where $(a_1,a_2,...,a_N)^T$ is the ground state as before. Then from
Eq.(26) and Eq.(28) we have
\begin{equation}
A_k=\frac{\sum_{i=1}^k{(\frac{1}{1+\frac{s}{t}f(i)})^2}}{\sum_{i=1}^N{(\frac{1}{1+\frac{s}{t}f(i)})^2}},
\ \ k=1,2,...,N,
\end{equation}
From Eq.(9) it holds that
\begin{equation}
\frac{H(s)}{1-s}=H_0+\frac{s}{1-s}H_1 \ (s<1).
\end{equation}
It can be seen that $\frac{\lambda(s)}{1-s}$, the ground state
eigenvalue of $\frac{H(s)}{1-s}$, is a strictly increasing function
of $s$ \cite{RB97}. So
\begin{equation}
\frac{d}{ds}(\frac{\lambda}{1-s})>0,
\end{equation}
which makes
\begin{equation}
\frac{d}{ds}(\frac{t}{1-s})=\frac{d}{ds}(\frac{1-s-\lambda}{1-s})<0.
\end{equation}
A simple calculation shows that
\begin{equation}
-\frac{dt}{ds}>\frac{t}{1-s},
\end{equation}
then
\begin{equation}
\frac{d}{ds}(\frac{s}{t})>\frac{s}{t}(\frac{1}{s}+\frac{1}{1-s}), \
\ 0<s<1.
\end{equation}
Calculating the derivative of Eq.(37) we obtain
\begin{align}
\frac{dA_k}{ds}=&2\sum_{i=2}^k\sum_{j=k+1}^N{a_i^2
a_j^2(\frac{f(j)\frac{d}{ds}(\frac{s}{t})}{1+\frac{s}{t}f(j)}-\frac{f(i)\frac{d}{ds}(\frac{s}{t})}{1+\frac{s}{t}f(i)})}
\cr
&+2a_1^2\sum_{i=k+1}^N{a_i^2\frac{f(i)\frac{d}{ds}(\frac{s}{t})}{1+\frac{s}{t}f(i)}}.
\end{align}
Since
\begin{equation}
\frac{f(j)\frac{d}{ds}(\frac{s}{t})}{1+\frac{s}{t}f(j)}>\frac{f(i)\frac{d}{ds}(\frac{s}{t})}{1+\frac{s}{t}f(i)}
\end{equation}
when $N\geq j>i\geq 1$, we have
\begin{equation}
\frac{dA_k}{ds}>2a_1^2\sum_{i=k+1}^N{a_i^2\frac{f(i)\frac{d}{ds}(\frac{s}{t})}{1+\frac{s}{t}f(i)}}.
\end{equation}
Let $m=\min_{2\leq i \leq N}f(i)$. If $m>1$, it holds that
\begin{eqnarray}
\aligned
\frac{f(i)\frac{d}{ds}(\frac{s}{t})}{1+\frac{s}{t}f(i)}>&\frac{\frac{s}{t}f(i)(\frac{1}{s}+\frac{1}{1-s})}{\frac{s}{t}f(i)+1}\\
>&\frac{\frac{f(i)}{t}+\frac{s}{t}f(i)}{1+\frac{s}{t}f(i)}\\
>&1.
\endaligned
\end{eqnarray}
Here, we use Eq.(42) and the fact $t<1<f(i)$ and $\frac{1}{1-s}>1$.
Similarly, if $m<1$, it follows that
\begin{eqnarray}
\aligned
\frac{f(i)\frac{d}{ds}(\frac{s}{t})}{1+\frac{s}{t}f(i)}>&\frac{\frac{s}{t}f(i)(\frac{1}{s}+\frac{1}{1-s})}{\frac{s}{t}f(i)+1}\\
=&m\frac{\frac{1}{m}\cdot\frac{f(i)}{t}+\frac{1}{m(1-s)}\frac{s}{t}f(i)}{1+\frac{s}{t}f(i)}\\
>&m.
\endaligned
\end{eqnarray}
Thus we obtain
\begin{equation}
\frac{f(i)\frac{d}{ds}(\frac{s}{t})}{1+\frac{s}{t}f(i)}>c,
\end{equation}
where $c=min\{m,1\}$. Substituting Eq.(48) into Eq.(45), we have
\begin{equation}
\frac{dA_k}{ds}>2c\cdot a_1^2\sum_{i=k+1}^N{a_i^2}.
\end{equation}
Note that
\begin{equation}
a_1^2>\frac{\sum_{i=1}^k{a_i^2}}{k}=\frac{1}{k}A_k.
\end{equation}
We finally obtain
\begin{equation}
\frac{dA_k}{ds}>\frac{2c}{k}A_k(1-A_k).
\end{equation}

According to quantum adiabatic theorem, we know that for any
positive $\delta$ we have a finite running time $T$ such that
\begin{equation}
|\langle \psi'(s)|\psi(s)\rangle|\geq1-\delta^2/2
\end{equation}
for any $s\in (0,1)$. Since
\begin{eqnarray}
\aligned
||\psi\rangle-|\psi'\rangle|^2=&2-2\langle\psi(s)|\psi'(s)\rangle\\
<&\delta^2,
\endaligned
\end{eqnarray}
it can be seen that for any $s$
\begin{equation}
\sum_{i=1}^k|a_i-b_i|^2<\delta^2.
\end{equation}
Here, we choose the global phase of $|\psi'(s)\rangle$ such that
$\langle\psi(s)|\psi'(s)\rangle$ is real. According to Cauchy's
inequality, it holds that
\begin{equation}
\sum_{i=1}^k|a_i-b_i|<\sqrt{k}\delta.
\end{equation}
Note that
\begin{equation}
|a_i+b_i|<2,
\end{equation}
it follows that
\begin{equation}
\sum_{i=1}^k|a_i^2-b_i^2|=\sum_{i=1}^k|a_i-b_i|\cdot
|a_i+b_i|<2\sqrt{k}\delta.
\end{equation}
Thus
\begin{equation}
|\sum_{i=1}^k|a_i|^2-\sum_{i=1}^k|b_i|^2|\leq\sum_{i=1}^k|a_i^2-b_i^2|<2\sqrt{k}\delta.
\end{equation}

Now let us consider two points $(s_1,A_k')$ and $(s_2,A_k)$ on
$s-A_k$ curve (about the ground state) and two points $(s_1,B_k')$
and $(s_2,B_k)$ on $s-B_k$ curve (about the actual state), where
$s_2-s_1=\triangle s$, $0<\triangle s\ll 1$. These four points are
all near the end of the quantum adiabatic evolution. If step-by-step
majorization of the actual state holds, $s-B_k$ curve should be a
monotonically increasing curve. Suppose that Eq.(52) holds.
According to Eq.(58) we have
\begin{equation}
|A_k-B_k|<2\sqrt{k}\delta, \ \ \ |A_k'-B_k'|<2\sqrt{k}\delta.
\end{equation}
On the other hand, according to the discussion above, it holds that
\begin{equation}
A_k-A_k' > \frac{2c}{k}A_k(1-A_k)\cdot\bigtriangleup s.
\end{equation}
Then if
\begin{equation}
\delta<\frac{c}{2k\sqrt{k}}A_k(1-A_k)\cdot\bigtriangleup s,
\end{equation}
we have $B_k'<B_k$.

Note that for arbitrary small $\bigtriangleup s$ we can find a
corresponding $\delta$ or running time $T$ such that Eq.(61) holds.
Thus, it can be judged that when the running time continues becoming
longer, the oscillation at the end of $s-B_k$ curve becomes weaker
and weaker. This explains the results of numerical simulations for
global adiabatic search algorithms in \cite{OLM021}, which is a
special case of our discussion above. In fact, this is consistent
with our intuition. By quantum adiabatic theorem, we know that when
the running time becomes longer, the distance between the actual
state of the system and the ground state becomes smaller. Since it
has been shown that the ground states shows exact step-by-step
majorization, it's natural that the actual state enjoys the same
relation approximately. The longer the running time, the better the
approximation.

It should be pointed out that this paper only deals with a special
class of quantum adiabatic algorithms. Whether all quantum adiabatic
algorithms enjoy step-by-step majorization (exactly or
approximately) remains open.

\section{CONCLUSION}

In conclusion, we have shown that for any algorithm of a special
class of quantum adiabatic algorithms, step-by-step majorization of
the ground state holds perfectly. We have also shown that
step-by-step majorization of the actual state holds approximately.
This supports the conclusion that majorization seems to appear
universally in quantum adiabatic algorithms. For further studies,
whether step-by-step majorization holds for more quantum adiabatic
algorithms should be examined carefully.

As mentioned at the beginning of this paper, step-by-step
majorization has been applied to analyze the efficiency of quantum
algorithms by Latorre et al \cite{LM02}. They pointed out that that
just obeying step-by-step majorization can not guarantee the
efficiency \cite{OLM021}. The results obtained in this paper offer
facts to indicate the same conclusion. It have been shown that the
running time of the class of algorithms discussed in this paper is
exponential in $n$, the problem size \cite{ZH06,FGGN05,RAO05,WY061}.
It seems that the performance of these algorithms are not very good.
However, these algorithms are in different situation in efficiency.
Some of them are optimal, such as local adiabatic search algorithm
\cite{RC02}, while the others are not, such as the adiabatic
algorithm for the hidden subgroup problem \cite{RAO03}. However, as
pointed out above, these algorithms enjoy the similar majorization
relation. On the other hand, Latorre et al illustrated that all
known fast and efficient quantum algorithms show step-by-step
majorization. For further studies, whether step-by-step majorization
is really necessary for efficiency is an important and interesting
question.

\section{ACKNOWLEDGMENTS}

We are grateful to R. Or\'{u}s, J. I. Latorre and M. A.
Mart\'{\i}n-Delgado for helpful comments and suggestions. We thank
the colleagues in the Quantum Computation and Information Research
Group for helpful discussions. This work was partly supported by the
National Nature Science Foundation of China (Grant Nos. 60503001,
60321002, and 60305005).

\end{document}